\documentclass[preprint,showpacs,preprintnumbers,amsmath,amssymb,superscriptaddress]{revtex4}

\usepackage{graphicx}
\usepackage{dcolumn}
\usepackage{bm}

\begin{document}

\preprint{APS/}

\title{An algorithm for decoherence analyses of lights through three-dimensional periodic microstructures}

\author{I. L. Ho$^*$}
\affiliation{Research Center for Applied Sciences, Academia Sinica, Taipei, Taiwan, R.O.C.}
\email{sunta.ho@msa.hinet.net}
\author{Y. C. Chang}
\affiliation{Research Center for Applied Sciences, Academia Sinica, Taipei, Taiwan, R.O.C.}
\author{W. Y. Li}
\affiliation{Institute of Electro-Optical Science and Engineering, National Cheng Kung University,
Tainan, Taiwan, R.O.C.}
\author{M. T. Lee}
\affiliation{Research Center for Applied Sciences, Academia Sinica, Taipei, Taiwan, R.O.C.}
\author{C. Y. Yin}
\affiliation{Research Center for Applied Sciences, Academia Sinica, Taipei, Taiwan, R.O.C.}

\date{\today}

\begin{abstract}
A transfer-matrix algorithm is presented herein as a beginning to study the transmission characteristics of
coherent light through three-dimensional periodic microstructures, in which the structures are
treated as two-dimensional-layer stacks and multiple reflections are considered negligible.
The spatial-correlated noise is further introduced layer by layer to realize the actual decoherence
of the light and allows for statistical investigation of the partial spatially coherent optics in
transparent mediums. Numerical analyses show comparable results to the Gaussian Schell model in
free-space cases, indicating the validity of the algorithms.
\end{abstract}

\maketitle
\section{Introduction}
Nanoscale structures have achieved novel functions in electro-optic devices such as optical
filters, optical modulators, phase conjugated systems, optical attenuators, beam amplifiers,
tunable lasers, holographic data storage and even as parts for optical logic systems
\cite{ap1,ap2,ap3,ap4,ap5,ap6,ap7,ap8,ap9,ap10,ap11,ap12,ap13,ap14} over the last few
decades. In contrast to the assumption of the ideal coherent (plane-wave) light in most theoretical studies
\cite{op1,op2,op3,op4,op5,op6,op7}, the associated decoherence characteristics and
intensity distribution of the light, e.g. by light-emitting diodes (LED) \cite{led1,led2,led3}, otherwise introduce
another critical issue for both fundamental research and potential industrial applications. Motivated by these concerns and
based on our previous study on coupled-wave theory \cite{op6}, we present a transfer-matrix algorithm as a beginning to study the propagation of
 coherent light through three-dimensional periodic microstructures. This work discretizes the structures into
two-dimensional-layer stacks composed of isotropic or birefringent materials in an arbitrary order,
and multiple reflections are considered negligible. By employing a concept similar to
the Langevin dynamics in the finite difference method \cite{lang1}, a noise phase characterized with spatial correlations
$\tau$ is further included layer by layer to mimic the actual decoherence of the light and allows for statistical
investigation of the partial spatially coherent optics. Here, the correlation length $\tau$ of the noise is assumed to be
larger than the wavelength of the incidence $\lambda$, because the intense decoherence (small $\tau$) causes strong
scattering and the multiple reflections thereby cannot be ignored. The generation of the noise function is fulfilled
by using a standard numerical technique based on the Fourier transform \cite{fft1,fft2} and is formulated in the appendix for the
two-dimensional cases studied. Numerical analyses for the propagation of the free-space Gaussian beam and the diffractions
of the beam incident through a liquid-crystal grating at different spatial decoherence show comparable results
to the Gaussian Schell model \cite{PCB1,PCB2} and can validate this study.

\section{Theoretical formulae}
This section neglects the multiple reflections and derives the transfer-matrix algorithm that is much easier
to manipulate algebraically, yet accounts for the effects of the Fresnel refraction and the single reflection
at the interfaces of the medium. The assumption of no multiple reflection is legitimate for most practical
transparent materials. In what follows, a reformulation, including the noise phase of electromagnetic fields,
 is fulfilled to demonstrate the decoherence behaviors of the light.
\subsection{Transfer matrix method by RCWA}
\begin{figure}[tbp]
\begin{center}
\includegraphics[scale=0.3]{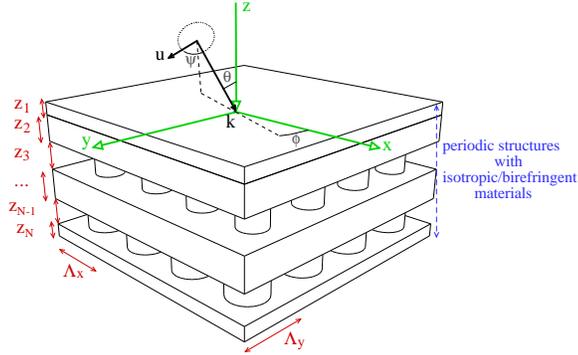}
\end{center}
\caption{Geometry of three-dimensional RCWA algorithm for a multi-layer stack with two-dimensional
periodic microstructures in arbitrary arrangement with isotropic and birefringent materials .} \label{str1}
\label{dynamics}
\end{figure}
Referring to Figure \ref{str1}, the structure is stratified along the medium normal into $N$ layers and these
layers contribute to a total thickness of the stacks $Z_{N}=\sum_{\ell=1}^{N}z_{\ell}$. Each layer is considered as
an arbitrary arrangement with homogeneous or birefringent materials, and the dielectric coefficients are generally
described by a matrix:
\begin{equation}
\varepsilon=\left[
\begin{array}{ccc}
\varepsilon _{xx} & \varepsilon _{xy} & \varepsilon _{xz} \\
\varepsilon _{yx} & \varepsilon _{yy} & \varepsilon _{yz} \\
\varepsilon _{zx} & \varepsilon _{zy} & \varepsilon _{zz}%
\end{array}%
\right]  \label{dimatrix}
\end{equation}
with
\begin{eqnarray}
\varepsilon _{xx} &=&n_{o}^{2}+\left( n_{e}^{2}-n_{o}^{2}\right) \sin
^{2}\theta _{o}\cos ^{2}\phi _{o},  \nonumber \\
\varepsilon _{xy} &=&\varepsilon _{yx}=\left( n_{e}^{2}-n_{o}^{2}\right)
\sin ^{2}\theta _{o}\sin \phi _{o}\cos \phi _{o},  \nonumber \\
\varepsilon _{xz} &=&\varepsilon _{zx}=\left( n_{e}^{2}-n_{o}^{2}\right)
\sin \theta _{o}\cos \theta _{o}\cos \phi _{o},  \nonumber \\
\varepsilon _{yy} &=&n_{o}^{2}+\left( n_{e}^{2}-n_{o}^{2}\right) \sin
^{2}\theta _{o}\sin ^{2}\phi _{o},  \nonumber \\
\varepsilon _{yz} &=&\varepsilon _{zy}=\left( n_{e}^{2}-n_{o}^{2}\right)
\sin \theta _{o}\cos \theta _{o}\sin \phi _{o},  \nonumber \\
\varepsilon _{zz} &=&n_{o}^{2}+\left( n_{e}^{2}-n_{o}^{2}\right) \cos
^{2}\theta _{o},
\end{eqnarray}%
Here, $n_{e}$ and $n_{o}$ are extraordinary and ordinary indices of
refraction of the uniaxially birefringent medium, respectively, $\theta _{o}$
is the angle between the optic axes and the $z$ axis, and $\phi _{o}$ is the
angle between the projection of the optic axes on the $xy$ plane and $x$ axis.

To introduce the rigorous coupled-wave theory to the stack, all of the layers are defined as having the same
periodicity: $\Lambda_{x}$ along the x direction and $\Lambda_{y}$ along the
y direction. The periodic permittivity of an individual layer $\ell$ in
the stack thereby can be expanded in the Fourier series of the spatial harmonics as:
{\small
\begin{eqnarray}
\overline{\varepsilon }_{ij}\left( \overline{x},\overline{y},\overline{z}%
_{\ell }\right) &=&\sum_{g,h}\overline{\varepsilon }_{ij,gh}\left( \overline{%
z}_{\ell }\right) \exp \left( i\frac{g\lambda \overline{x}}{\Lambda _{x}}+i%
\frac{h\lambda \overline{y}}{\Lambda _{y}}\right)  \label{keps} \\
\overline{\varepsilon }_{ij,gh}\left( \overline{z}_{\ell }\right) &=&\frac{%
\lambda }{2\pi \Lambda _{x}}\frac{\lambda }{2\pi \Lambda _{y}}\int_{0}^{%
\frac{2\pi \Lambda _{x}}{\lambda }}\int_{0}^{\frac{2\pi \Lambda _{y}}{%
\lambda }}\overline{\varepsilon }_{ij}\left( \overline{x},\overline{y},%
\overline{z}_{\ell }\right) \exp \left( -i\frac{g\lambda \overline{x}}{%
\Lambda _{x}}-i\frac{h\lambda \overline{y}}{\Lambda _{y}}\right) d\overline{x%
}d\overline{y}  \label{kepsi}
\end{eqnarray}%
}Here, we have defined variables $k_{0}=\omega \sqrt{\mu _{0}\varepsilon _{0}}=\frac{2\pi }{%
\lambda }$, $Y_{0}=\frac{1}{Z_{0}}=\sqrt{\frac{\varepsilon _{0}}{\mu _{0}}}$%
, $\overline{r}=k_{0}r$, $\overline{x}=k_{0}x$, $\overline{y}=k_{0}y$, and $%
\overline{z}=k_{0}z$. $\lambda$ is the vacuum wavelength of the incident wave.
Note that $\varepsilon_{ij\in \{x,y,z\}}$ are defined as functions of position ($x,y,z$)
and $\overline{\varepsilon}_{ij}$ are defined for ($\overline{x},\overline{y},\overline{z}$).
A parallel transform for the electromagnetic fields through the stack is expressed in terms
of Rayleigh expansions: {\small
\begin{eqnarray}
\sqrt{Y_{0}}\mathbf{E}\left( \overline{x},\overline{y},\overline{z}_{\ell
}\right) &=&\sum_{g,h}\mathbf{e}_{gh}\left( \overline{z}_{\ell }\right) \exp %
\left[ -i\left( n_{xg}\overline{x}+n_{yh}\overline{y}\right) \right]
\label{ke} \\
\sqrt{Z_{0}}\mathbf{H}\left( \overline{x},\overline{y},\overline{z}_{\ell
}\right) &=&\sum_{g,h}\mathbf{h}_{gh}\left( \overline{z}_{\ell }\right) \exp %
\left[ -i\left( n_{xg}\overline{x}+n_{yh}\overline{y}\right) \right]
\label{kh} \\
n_{xg} &=&n_{I}\sin \theta \cos \phi -g\frac{\lambda }{\Lambda _{x}}
\label{diffg} \\
n_{yh} &=&n_{I}\sin \theta \sin \phi -h\frac{\lambda }{\Lambda _{y}}
\label{diffh}
\end{eqnarray}%
}This indicates the propagation of light along the direction
$\mathbf{n_{gh}}=n_{xg}\hat{\imath}+n_{yh}\hat{\jmath}+\xi_{gh}\hat{k}$ with
$\xi _{gh}=(n^{2}_{I(E)}-n_{yh}n_{yh}-n_{xg}n_{xg})^{1/2}$,
in which $n_{I}$ and $n_{E}$ are the refraction index and correspond to the propagation in
the incident and emitted regions, respectively. $\theta $, $\phi $ are the incident
angles defined by sphere coordinates, and $z$ is the normal direction for
the $xy$ plane of periodic structures.

Organizing the algorithms from our previous study \cite{op6} while ignoring
the multiple reflections,
the transfer-matrix formulae for the microstructure can thus be written as:
\begin{equation}
\left[
\begin{array}{c}
\vec{E}_{q,N+1}^{+} \\
\vec{M}_{q,N+1}^{+} \\
\vec{E}_{q,N+1}^{-} \\
\vec{M}_{q,N+1}^{-}%
\end{array}%
\right] =\mathbf{M}_{ext}\mathbf{M}_{N}...\mathbf{M}_{2}\mathbf{M}_{1}\mathbf{M}_{ent}\left[
\begin{array}{c}
\vec{E}_{q,0}^{+} \\
\vec{M}_{q,0}^{+} \\
\vec{E}_{q,0}^{-} \\
\vec{M}_{q,0}^{-}%
\end{array}%
\right] \label{TM}
\end{equation}
Here, $\mathbf{M}_{i\in\{1\sim N\}}$ is the transfer matrix corresponding to the $i_{th}$ (anisotropic)
structured layer. It is formulated by the
eigen-vector matrix $\mathbf{T}^{(a)}_{i}$ with column eigen-vectors and the diagonal eigen-value matrix
$\mathbf{\kappa}^{(a)}_{i}$ of the characteristic matrix $\mathbf{G}_{i}$ by:
{\small
\begin{eqnarray}
\mathbf{M}_{i} &=&\mathbf{T}_{i}^{(a)} exp \left[ i\mathbf{\kappa }_{i}^{a}%
\overline{z}_{i}\right] ( \mathbf{T}_{i}^{(a)}) ^{-1} \\
\mathbf{G}_{i} &=&\left[
\begin{array}{cccc}
\tilde{n}_{x}\tilde{\varepsilon}_{zz}^{-1}\tilde{\varepsilon}_{zx} & \tilde{n%
}_{x}\tilde{\varepsilon}_{zz}^{-1}\tilde{n}_{x}-1 & \tilde{n}_{x}\tilde{%
\varepsilon}_{zz}^{-1}\tilde{\varepsilon}_{zy} & -\tilde{n}_{x}\tilde{%
\varepsilon}_{zz}^{-1}\tilde{n}_{y} \\
\tilde{\varepsilon}_{xz}\tilde{\varepsilon}_{zz}^{-1}\tilde{\varepsilon}%
_{zx}-\tilde{\varepsilon}_{xx}+\tilde{n}_{y}\tilde{n}_{y} & \tilde{%
\varepsilon}_{xz}\tilde{\varepsilon}_{zz}^{-1}\tilde{n}_{x} & \tilde{%
\varepsilon}_{xz}\tilde{\varepsilon}_{zz}^{-1}\tilde{\varepsilon}_{zy}-%
\tilde{\varepsilon}_{xy}-\tilde{n}_{y}\tilde{n}_{x} & -\tilde{\varepsilon}%
_{xz}\tilde{\varepsilon}_{zz}^{-1}\tilde{n}_{y} \\
\tilde{n}_{y}\tilde{\varepsilon}_{zz}^{-1}\tilde{\varepsilon}_{zx} & \tilde{n%
}_{y}\tilde{\varepsilon}_{zz}^{-1}\tilde{n}_{x} & \tilde{n}_{y}\tilde{%
\varepsilon}_{zz}^{-1}\tilde{\varepsilon}_{zy} & -\tilde{n}_{y}\tilde{%
\varepsilon}_{zz}^{-1}\tilde{n}_{y}+1 \\
-\tilde{\varepsilon}_{yz}\tilde{\varepsilon}_{zz}^{-1}\tilde{\varepsilon}%
_{zx}+\tilde{\varepsilon}_{yx}+\tilde{n}_{x}\tilde{n}_{y} & -\tilde{%
\varepsilon}_{yz}\tilde{\varepsilon}_{zz}^{-1}\tilde{n}_{x} & -\tilde{%
\varepsilon}_{yz}\tilde{\varepsilon}_{zz}^{-1}\tilde{\varepsilon}_{zy}+%
\tilde{\varepsilon}_{yy}-\tilde{n}_{x}\tilde{n}_{x} & \tilde{\varepsilon}%
_{yz}\tilde{\varepsilon}_{zz}^{-1}\tilde{n}_{y}%
\end{array}%
\right]
\end{eqnarray}%
}where the notation $\vec{(.)}$ represents $N_{g}\times N_{h}$ vectors with components $(.)_{gh}$, $\tilde{n}_{x}$
($\tilde{n}_{y}$) are $N_{g}N_{h}\times N_{g}N_{h}$ diagonal matrices with the diagonal elements $n_{xg}$ ($n_{yh}$),
and $\tilde{\varepsilon}_{ij\in\{x,y,z\}}$ are $N_{g}N_{h}\times N_{g}N_{h}$ matrices with elements
$\varepsilon_{ij,gh}$ formulated by $\vec{f}\sim \tilde{\varepsilon}_{ij}\vec{f^{\prime}} $ or
$f_{gh}\sim \sum_{g^{\prime }h^{\prime }}\varepsilon _{ij,(g-g^{\prime})(h-h^{\prime })}f^{\prime}_{g^{\prime }h^{\prime }}$.
$N_{g(h)}$ define the number of considered Fourier orders $g$ ($h$) in the $x$ ($y$) direction.
$1$ is $N_{g}N_{h}\times N_{g}N_{h}$ identity matrix.
One may understood that $( \mathbf{T}_{i}^{(a)} )^{-1}$ term represents the coordinate
transformation from the components of tangential fields
$\mathbf{f}_{\hat{t},i}=[\vec{e}_{x,i} \ \vec{h}_{y,i} \ \vec{e}_{y,i} \ \vec{h}_{x,i}]^{t}$ at
$i_{th}$ interface in Equations (\ref{ke})-(\ref{kh})
into the independent components of the eigen-modes of the $i_{th}$ layer,
$exp \left[ i\mathbf{\kappa }_{i}^{a}\overline{z}_{i}\right]$ term then follows the corresponding
eigen-mode transition lasting a distance $\overline{z}_{i}$, and $\mathbf{T}_{i}^{(a)}$ term is the
inversely coordinate transformation back to the components of the tangential fields defining
$\mathbf{f}_{\hat{t},i+1}$ at the next interface. The superscript $t$ indicates the matrix transposition.
Summarily, $\mathbf{M}_{i}$ demonstrates the propagation of fields ($\mathbf{f}_{\hat{t},i}$ to
$\mathbf{f}_{\hat{t},i+1}$) through the $i_{th}$ layer.

In the (isotropic) uniform incident ($i=0$) and emitted ($i=N+1$) regions, the eigen-modes are chosen as
$\vec{E}_{q}^{+}$ and $\vec{M}_{q}^{+}$
($\vec{E}_{q}^{-}$ and $\vec{M}_{q}^{-}$) \cite{op6,op7}, thereby representing the physical forward
(backward) $TE$ and $TM$ waves, i.e. transverse electric and transverse magnetic fields
corresponding to the plane of the diffraction wave, respectively. The transform between the eigen-modes and
the tangential fields $\mathbf{f}_{\hat{t},0}=[\vec{e}_{x,0} \ \vec{h}_{y,0} \ \vec{e}_{y,0} \ \vec{h}_{x,0}]^{t}$
in the incident region ($i=0$) is written as:
{\small
\begin{eqnarray}
\left[
\begin{array}{c}
\vec{e}_{x,0} \\
\vec{h}_{y,0} \\
\vec{e}_{y,0} \\
\vec{h}_{x,0}%
\end{array}%
\right]  &=&\left[
\begin{array}{cccc}
\dot{\mathbf{n}}_{y} & \dot{\mathbf{n}}_{x} & \dot{\mathbf{n}}_{y} & \dot{%
\mathbf{n}}_{x} \\
\dot{\mathbf{n}}_{y}\mathbf{\xi } & \mathbf{\varepsilon_{I} \dot{\mathbf{n}}%
_{x}\xi }^{-1} & -\dot{\mathbf{n}}_{y}\mathbf{\xi } & -\mathbf{\varepsilon_{I}
\dot{\mathbf{n}}_{x}\xi }^{-1} \\
-\dot{\mathbf{n}}_{x} & \dot{\mathbf{n}}_{y} & -\dot{\mathbf{n}}_{x} & \dot{%
\mathbf{n}}_{y} \\
\dot{\mathbf{n}}_{x}\mathbf{\xi } & -\mathbf{\varepsilon_{I} \dot{\mathbf{n}}%
_{y}\xi }^{-1} & -\dot{\mathbf{n}}_{x}\mathbf{\xi } & \mathbf{\varepsilon_{I}
\dot{\mathbf{n}}_{y}\xi }^{-1}%
\end{array}%
\right] \left[
\begin{array}{c}
\vec{E}_{q,0}^{+} \\
\vec{M}_{q,0}^{+} \\
\vec{E}_{q,0}^{-} \\
\vec{M}_{q,0}^{-}%
\end{array}%
\right]   \nonumber \\
&\equiv&\mathbf{T}^{(i)}_{\varepsilon_{I}}\left[
\begin{array}{c}
\vec{E}_{q,0}^{+} \\
\vec{M}_{q,0}^{+} \\
\vec{E}_{q,0}^{-} \\
\vec{M}_{q,0}^{-}%
\end{array}%
\right] \label{isoT}
\end{eqnarray}}Here, $\dot{\mathbf{n}}_{y}$ and $\dot{\mathbf{n}}_{x}$ are $%
N_{g}N_{h}\times N_{g}N_{h}$ diagonal matrices with normalized elements $%
\frac{n_{yh}}{m_{gh}}$ and $\frac{n_{xg}}{m_{gh}}$ respectively. $\mathbf{%
\xi }^{-1}$ is the matrix with elements $1/\xi _{gh}$ (not the inverse of
the matrix $\mathbf{\xi }$), in which $m_{gh}=(n_{yh}n_{yh}+n_{xg}n_{xg})^{1/2}$,
$\xi _{gh}=(\varepsilon_{I}-n_{yh}n_{yh}-n_{xg}n_{xg})^{1/2}$, and $\varepsilon_{I}=n_{I}^{2}$
have been defined for the incident region.
A similar transform for $\mathbf{f}_{\hat{t},N+1}$ in the emitted region can be derived
straightforwardly by replacing all the
$\varepsilon_{I}$ in Equation (\ref{isoT}) with $\varepsilon_{E}$ and can be obtained as
$\mathbf{f}_{\hat{t},N+1}=\mathbf{T}^{(i)}_{\varepsilon_{E}}[\vec{E}_{q,N+1}^{+} \ \vec{M}_{q,N+1}^{+} \
\vec{E}_{q,N+1}^{-} \ \vec{M}_{q,N+1}^{-}]^{t} $, in which
$\xi _{gh}=(\varepsilon_{E}-n_{yh}n_{yh}-n_{xg}n_{xg})^{1/2}$, and $\varepsilon_{E}=n_{E}^{2}$.

Moreover, $\mathbf{M}_{ent}$ is the matrix representing the light propagation from
the incident region into the medium, and indicates the essential refraction and the
reflection at the first interface of the medium. To consider these effects in a simple way,
a virtual (isotropic) uniform layer, which has zero thickness and (scalar) average
dielectric coefficient $\varepsilon_{a}=n^{2}_{avg}$, e.g. $n_{avg}=(n_{e}+n_{o})/2$
for the liquid-crystal grating, is assumed to exist between the incident region and the $1_{st}$
layer. $\mathbf{M}_{ent}$ thereby can be obtained as:
\begin{eqnarray}
\mathbf{M}_{ent} &=&\mathbf{T}_{\mathbf{\varepsilon }_{a}}^{(i)}\left[
\begin{array}{cc}
\mathbf{W}_{1}^{-1} & \mathbf{0} \\
\mathbf{0} & \mathbf{0}%
\end{array}%
\right]  \\
\left[
\begin{array}{cc}
\mathbf{W}_{1} & \mathbf{W}_{2} \\
\mathbf{W}_{3} & \mathbf{W}_{4}%
\end{array}%
\right]  &=&\left[ (\mathbf{T}_{\mathbf{\varepsilon }_{a}}^{(i)})^{-1}%
\mathbf{T}_{\mathbf{\varepsilon }_{I}}^{(i)}\right] ^{-1}
\end{eqnarray}
Here, $\mathbf{T}_{\mathbf{\varepsilon }_{a}}^{(i)}$ is defined in Equation
(\ref{isoT}) with the replacements of $\varepsilon_{I}$ by $\varepsilon_{a}$,
$\xi _{gh}=(\varepsilon_{a}-n_{yh}n_{yh}-n_{xg}n_{xg})^{1/2}$, and $\varepsilon_{a}=n_{avg}^{2}$.
Parallel to the argument of $\mathbf{M}_{ent}$, another similar virtual (isotropic) uniform layer
exists between the emitted region and the $N_{st}$ layer, and $\mathbf{M}_{ext}$ is approximated as:
\begin{eqnarray}
\mathbf{M}_{ext} &=&\left[
\begin{array}{cc}
\mathbf{W}_{1}^{\prime -1} & \mathbf{0} \\
\mathbf{0} & \mathbf{0}%
\end{array}%
\right] (\mathbf{T}_{\mathbf{\varepsilon }_{a}}^{(i)})^{-1} \\
\left[
\begin{array}{cc}
\mathbf{W}_{1}^{\prime } & \mathbf{W}_{2}^{\prime } \\
\mathbf{W}_{3}^{\prime } & \mathbf{W}_{4}^{\prime }%
\end{array}%
\right]  &=&\left[ (\mathbf{T}_{\mathbf{\varepsilon }_{E}}^{(i)})^{-1}%
\mathbf{T}_{\mathbf{\varepsilon }_{a}}^{(i)}\right] ^{-1}
\end{eqnarray}

\subsection{Algorithms extended for optical decoherences}
To demonstrate the spatial decoherence of the light, the electric fields of an incident
unit-amplitude plane wave are first denoted as:
\begin{equation}
\mathbf{E}\left( \mathbf{r}\right) =\mathbf{E}_{0}\exp \left[ -i\mathbf{k}%
\cdot \mathbf{r}+i\eta \left( \mathbf{r}\right) \right]  \label{efield}
\end{equation}%
The time dependence of $\exp \left( i\omega t\right) $ is assumed and omitted
here. $\eta (x,y,z)$ is the noise phase of $\mathbf{E}$ and is conditioned by the correlation relation with
a characteristic spatial coherent length $\tau $:
\begin{eqnarray}
\left\langle \eta \left( \mathbf{r}\right) \eta \left( \mathbf{r}^{\prime
}\right) \right\rangle &=&\gamma \left( \mathbf{r-r}^{\prime }\right)
\nonumber  \label{corr} \\
&=&\pi ^{2}\exp \left( -\frac{\left\vert \mathbf{r-r}^{\prime }\right\vert
^{2}}{2\tau ^{2}}\right)
\end{eqnarray}%
with $\left\langle \eta \left( \mathbf{r}\right) \right\rangle=0$. The correlation function
$\gamma \left( \mathbf{r-r}%
^{\prime }\right) $ is chosen as Gaussian distribution and describes a strong (weak) phase
correlation between fields of closer (farther) positions than $\tau $. In
the limit of $\tau \rightarrow \infty $, the $\eta\left( \mathbf{r} \right)$
is constant and the $\mathbf{E}$ field in Equation(%
\ref{efield}) shows the coherent incident wave, whereas in the limit of $%
\tau \rightarrow 0$, an incoherent (total random phase) wave is exhibited.
Here, the generation of the noise function $\eta \left( \mathbf{r}\right)$ is fulfilled by using a
standard numerical technique based on the Fourier transform and is formulated in the appendix for the
two-dimensional cases studied.
To realize the spatial decoherence analyses for the light, we introduce the concept similar to the
Langevin dynamics for the stratified layers along the $z$ direction and include the noise phase layer by layer in
Equation (\ref{TM}):
\begin{equation}
\left[
\begin{array}{c}
\vec{E}_{q,N+1}^{+} \\
\vec{M}_{q,N+1}^{+} \\
\vec{E}_{q,N+1}^{-} \\
\vec{M}_{q,N+1}^{-}%
\end{array}%
\right] =\mathbf{M}_{ext}\grave{\mathbf{R}}\mathbf{M}_{N}...\grave{\mathbf{R}%
}\mathbf{M}_{2}\grave{\mathbf{R}}\mathbf{M}_{1}\grave{\mathbf{R}}\mathbf{M}%
_{ent}\left[
\begin{array}{c}
\vec{E}_{q,0}^{+} \\
\vec{M}_{q,0}^{+} \\
\vec{E}_{q,0}^{-} \\
\vec{M}_{q,0}^{-}%
\end{array}%
\right]   \label{TMR}
\end{equation}
Here, $\grave{\mathbf{R}}$ indicates the operations of three numerical processes
on
$\mathbf{f}_{\hat{t},i\in\{1\sim N+1\}}=\mathbf{M}_{i-1}...\grave{\mathbf{R}}\mathbf{M}_{2}\grave{\mathbf{R}}\mathbf{M}_{1}\grave{\mathbf{R}}\mathbf{M}%
_{ent} \cdot[
\vec{E}_{q,0}^{+} \
\vec{M}_{q,0}^{+} \
\vec{E}_{q,0}^{-} \
\vec{M}_{q,0}^{-}%
 ]^{t}$  at $i_{th}$ interface by: (a) first fulfilling the inverse Fourier transform from the $k$ components
 of the tangential fields
$\mathbf{f}_{\hat{t},i}(\bar{x},\bar{y})$ into the $r$ components, i.e.
($\sqrt{Y_{0}}\mathbf{E}$, $\sqrt{Z_{0}}\mathbf{H}$ ) by Equation (\ref{ke}); (b) next including the noise phase by
multiplying the spatial $r$ components of the fields by the noise phase $e^{i\eta (x,y,z)}$, or
$e^{i\eta (\bar{x},\bar{y},\bar{z})}$ in our work, as Equation (\ref{efield});
(c) and finally performing the Fourier transform on the $r$ components to obtain the
$k$ components of the tangential fields
$\mathbf{f}_{\hat{t},i}(\bar{x},\bar{y})$ including the noise phase.
Hence, the field $\mathbf{f}_{\hat{t},i\in\{1\sim N+1\}}$ through the medium can approximately
mimic decoherence behaviors by these processes.

\section{Numerical results for two-dimensional examples }
\begin{figure}[tbp]
\begin{center}
\includegraphics[scale=0.75]{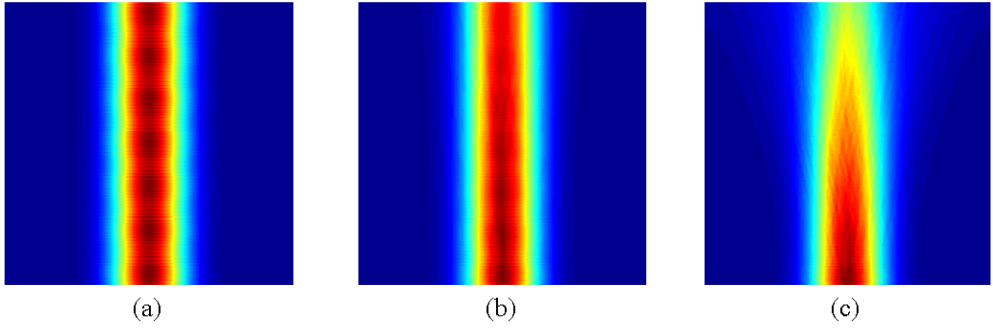}
\end{center}
\caption{Intensity patterns of the free-space Gaussian beam with coherent length
(a) $\tau=5 \mu m$, (b) $\tau=1 \mu m$, and (c) $\tau=0.5 \mu m$ for the ensemble
of 500 iterations in the $48\mu m\times 40\mu m$ $xz$ plane.} \label{ex1field}
\end{figure}

\begin{figure}[tbp]
\begin{center}
\includegraphics[scale=0.65]{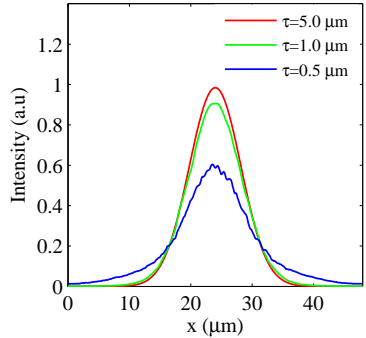}
\end{center}
\caption{Numerical intensity of the Gaussian beam corresponding to the results
of Figure \ref{ex1field} with $\tau=5 \mu m$, $\tau=1 \mu m$, and $\tau=0.5 \mu m$ after propagating
a distance at $z=35\mu m$.} \label{ex1ana}
\end{figure}

To study the decoherence behavior of light, we first analyze the propagation of the Gaussian beam with different
coherent lengths in a two-dimensional $xz$ free space, and fulfill a statistical numerical analysis
over an ensemble of systems with 500 identical iterations. The characteristic period of the structure
($\Lambda_{x}=48\mu m$) is chosen to be much larger than the Gaussian beam profile
($\sim e^{-x^{2}/2\sigma^{2}}$ with $\sigma=6\mu m$) so that the periodic beam can be considered to
be isolated in near-field optic analyses. The relevant parameters are given as: the x-grid size
$\Delta_{x}=0.05\mu m$, the thickness of the stacked layer $z_{i\in\{1\sim N\}}=0.05\mu m$,
the wavelength $\lambda=0.55\mu m$, the $y$-polarized incidence
$\sqrt{Y_{0}}E_{y}(\bar{x},\bar{z}=0)=e^{-\bar{x}^{2}/2\sigma^{2}}$, the refraction indices
$n_{I}=n_{E}=n_{o}=1.0\simeq n_{e}=1.0+10^{-5}$, and the Fourier order $-50<g<50$ ($h=0$).
Figure \ref{ex1field} illustrates the intensity patterns
$|\sqrt{Y_{0}}E_{y}(\bar{x},\bar{z})|^{2}+|\sqrt{Z_{0}}H_{x}(\bar{x},\bar{z})|^{2}$
of the free-space Gaussian beam with $\tau=5\mu m$, $\tau=1\mu m$, and $\tau=0.5\mu m$
along the light propagation. Figure \ref{ex1ana} shows the corresponding numerical intensity
to the Figure \ref{ex1field} after propagating a distance at $z=35 \mu m$. The results indicate
that the stronger decoherence (smaller $\tau$) of the light causes more a intense scattering along
the propagation and shows comparable conclusion from the Gaussian Schell model \cite{PCB1}.

\begin{figure}[tbp]
\begin{center}
\includegraphics[scale=0.75]{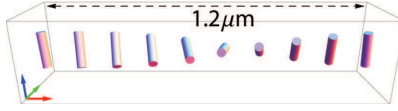}
\end{center}
\caption{One-period orientations of liquid-crystal directors in the single-layer film, in which the red,
green, and blue arrows indicate the spatial $\hat{x}$, $\hat{y}$,
and $\hat{z}$ directions, respectively.} \label{ex2str}
\end{figure}

\begin{figure}[tbp]
\begin{center}
\includegraphics[scale=0.75]{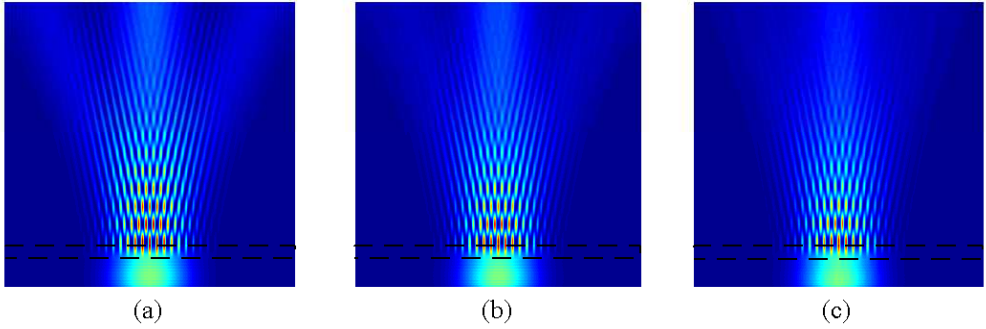}
\end{center}
\caption{Intensity patterns of the Gaussian beam incident through a forty-period grating (dashed-line rectangle)
in the $48\mu m\times 40\mu m$ $xz$ plane. The coherent lengths of the beam are statistically considered to
 be (a) $\tau=5 \mu m$, (b) $\tau=1 \mu m$, and (c) $\tau=0.5 \mu m$ in an ensemble of 500 iterations.}
 \label{ex2field}
\end{figure}

\begin{figure}[tbp]
\begin{center}
\includegraphics[scale=0.65]{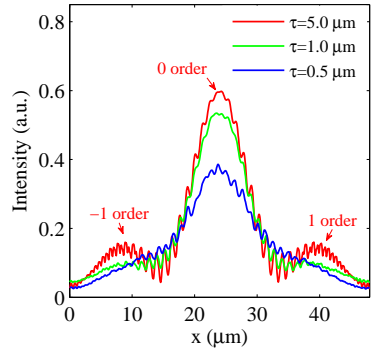}
\end{center}
\caption{Numerical intensity of the Gaussian beam corresponding to the results of Figure \ref{ex2field}
with $\tau=5 \mu m$, $\tau=1 \mu m$, and $\tau=0.5 \mu m$ after propagating a distance at $z=35\mu m$.}
\label{ex2ana}
\end{figure}

To further investigate the decoherence behaviors of the light through the microstructures,
a liquid-crystal grating film is included to introduce the diffraction of light.
Figure \ref{ex2str} shows the one-period orientations of the liquid-crystals directors in
a single-layer film. For this case, rather than employing the definition of the grating period
$\Lambda_{x,LC}=1.2\mu m$ as in Figure \ref{ex2str}, we apply SRC-RCWA scheme and set the
characteristic period $\Lambda_{x}=48\mu m=24\times\Lambda_{x,LC}$. Hence,
the long-range profile of the Gaussian beam in spatial space can be considered,
and simultaneously the small-angle emittance of the beam ($\sim 1/\Lambda_{x}$) defined by the
Fourier component in Equations (\ref{diffg})-(\ref{diffh}) can be demonstrated. Figure \ref{ex2field}
illustrates diffraction patterns
$|\sqrt{Y_{0}}E_{y}(\bar{x},\bar{z})|^{2}+|\sqrt{Z_{0}}H_{x}(\bar{x},\bar{z})|^{2}$ of the beam
incident through a forty-period liquid-crystal grating (enclosed by dashed lines) in
$48\mu m\times 40\mu m$ $xz$ plane, in which the coherent lengths of the beam are statistically
considered to be (a) $\tau=5 \mu m$, (b) $\tau=1 \mu m$, and (c) $\tau=0.5 \mu m$ for an ensemble
of 500 iterations. Figure \ref{ex2ana} shows the corresponding numerical intensity of the beam to
Figure \ref{ex2field} after propagating a distance at $z=35 \mu m$. The results indicate that
the coherent incidence ($\tau=5\mu m$) through gratings leads to definite diffractions as described
in most studies \cite{op6}, while the decoherent one ($\tau=0.5\mu m$) exhibits an intense scattering
similar to that in Figure \ref{ex1field}(c) and Figure \ref{ex1ana}. The incidence with $\tau=1\mu m$
 in Figure \ref{ex2field}(b) and Figure \ref{ex2ana} shows a simultaneous diffraction and
 scattering results and deviates from the results by the coherent optics.

\section{Conclusions}
This work has presented the formulas of the transfer-matrix method to conduct decoherence analyses in
three-dimensional periodic microstructures. The algorithms are also devoted to doing a per-study of the
light through turbid mediums that are related to the interaction between the fluctuated electron/dipole
 motions and the decoherence behaviors of light. Two numerical analyses for the propagation of free-space
  Gaussian beams and the diffraction of the liquid-crystal grating are then applied to verify the validity
  of this work, obtaining reasonable results.
\section{Acknowledgements}
This work was supported in part by the National Science Council of the Republic of China under Contract
Nos. NSC 99-2811-E-001-003 and NSC 98-2622-E-001-001-CC2.
\appendix
\section{Generation of two-dimensional spatial correlated noises}
In the following
context, we introduce the generation of fluctuation function $\eta \left(
\mathbf{r}\right) $ by the discrete Fourier method, and thereby realizing the $\mathbf{E}$
field describing partial coherence ($0<\tau <\infty $) in Equation (\ref{efield}%
). First, we introduce $\eta(\mathbf{k}) $ in $2D$ momentum space, which follows the
transforms of:

\begin{eqnarray}
\eta \left( \mathbf{k}\right) &=&\int_{-\infty }^{\infty }\int_{-\infty
}^{\infty }\eta \left( \mathbf{r}\right) \exp \left( -i\mathbf{k}\cdot
\mathbf{r}\right) d\mathbf{r} \\
\eta \left( \mathbf{r}\right) &=&\frac{1}{4\pi ^{2}}\int_{-\infty }^{\infty
}\int_{-\infty }^{\infty }\eta \left( \mathbf{k}\right) \exp \left( i\mathbf{%
k}\cdot \mathbf{r}\right) d\mathbf{k}
\end{eqnarray}%
or alternatively the discrete representation in $N\times N$ grid data under
spatial interval $\Delta $ for our case

\begin{eqnarray}
\eta _{g,h} &=&\Delta ^{2}\sum_{u,v=0}^{N-1}\eta _{u,v}\exp \left[ -i\frac{%
2\pi }{N}\left( g\cdot u+h\cdot v\right) \right] ,\text{ \ \ }g,h=0,1,...,N-1%
\text{\ \ } \label{disFT1} \\
\eta _{u,v} &=&\frac{1}{N^{2}\Delta ^{2}}\sum_{g,h=0}^{N-1}\eta _{g,h}\exp %
\left[ i\frac{2\pi }{N}\left( g\cdot u+h\cdot v\right) \right] ,\text{ \ \ }%
u,v=0,1,...,N-1 \label{disFT2}
\end{eqnarray}%
Here $u$, $v$\ ($g$,$h$) are the indices for the spatial (momentum) space. The
correlation function of the momentum space corresponding to that of the spatial space in Equation (\ref{corr}%
) is next derived as:
\begin{eqnarray}
\left\langle \eta \left( \mathbf{k}\right) \eta \left( \mathbf{k}^{\prime
}\right) \right\rangle &=&4\pi ^{2}\delta \left( \mathbf{k+k}^{\prime
}\right) \gamma \left( \mathbf{k}\right)  \nonumber  \label{kform} \\
&=&4\pi ^{2}\delta \left( \mathbf{k+k}^{\prime }\right) \cdot 2\pi
^{3}\tau ^{2}\exp \left[ -\frac{1}{2}\mathbf{k}^{2}\tau ^{2}\right]
\end{eqnarray}%
where $\gamma \left( \mathbf{k}\right) $ is the Fourier transform of $\gamma
\left( \mathbf{r}\right) $. The discrete representation of Equation (\ref{kform})
is
\begin{equation}
\left\langle \eta _{g,h}\eta _{g^{\prime },h^{\prime }}\right\rangle
=N^{2}\Delta ^{2}\gamma _{g,h}\delta _{g,N-g^{\prime }}\delta
_{h,N-h^{\prime }}  \label{kcorr}
\end{equation}%
Here, $\gamma _{g,h}$ is the discrete representation for $\gamma \left( \mathbf{k}%
\right) $ and exhibits the symmetry of $\gamma _{g,h}=\gamma _{N-g,h}=\gamma
_{g,N-h}=\gamma _{N-g,N-h}$. Accordingly, by the discrete correlation
equation in Equation (\ref{kcorr}), the fluctuation function $\eta _{g,h}$ can be
generated in the discrete momentum space by:

\begin{equation}
\eta _{g,h}=N\Delta \sqrt{\gamma _{g,h}}\alpha _{g,h}
\end{equation}%
in which $\alpha _{g,h}$ is the Gaussian random number (complex number) with the
average being zero, and it is conditioned as $\left\langle \alpha _{g,h}\alpha
_{g^{\prime },h^{\prime }}\right\rangle =\delta _{g,N-g^{\prime }}\delta
_{h,N-h^{\prime }}$, which corresponds to Equation (\ref{kcorr}).  Definitely, these
requirements for $\alpha _{g,h}$ can be realized by a simple process as:
\begin{eqnarray}
\alpha _{0,0} &=&a_{0,0},\text{ \ \ \ }\alpha _{\frac{N}{2},\frac{N}{2}}=a_{%
\frac{N}{2},\frac{N}{2}}  \nonumber \\
\alpha _{\frac{N}{2},0} &=&b_{\frac{N}{2},0},\text{ \ \ \ }\alpha _{0,\frac{N%
}{2}}=b_{0,\frac{N}{2}}  \nonumber \\
\alpha _{g,0} &=&\frac{1}{\sqrt{2}}\left( a_{g,0}+ib_{g,0}\right) ,\text{ \
\ \ }\alpha _{N-g,0}=\frac{1}{\sqrt{2}}\left( a_{g,0}-ib_{g,0}\right)
\nonumber \\
\alpha _{0,h} &=&\frac{1}{\sqrt{2}}\left( a_{0,h}+ib_{0,h}\right) ,\text{ \
\ \ }\alpha _{0,N-h}=\frac{1}{\sqrt{2}}\left( a_{0,h}-ib_{0,h}\right)
\nonumber \\
\alpha _{g,\frac{N}{2}} &=&\frac{1}{\sqrt{2}}(a_{g,\frac{N}{2}}+ib_{g,\frac{N%
}{2}}),\text{ \ \ \ }\alpha _{N-g,\frac{N}{2}}=\frac{1}{\sqrt{2}}(a_{g,\frac{%
N}{2}}-ib_{g,\frac{N}{2}})  \nonumber \\
\alpha _{\frac{N}{2},h} &=&\frac{1}{\sqrt{2}}(a_{\frac{N}{2},h}+ib_{\frac{N}{%
2},h}),\text{ \ \ \ }\alpha _{\frac{N}{2},N-h}=\frac{1}{\sqrt{2}}(a_{\frac{N%
}{2},h}-ib_{\frac{N}{2},h})  \nonumber \\
\alpha _{g,h} &=&\frac{1}{\sqrt{2}}\left( a_{g,h}+ib_{g,h}\right) ,\text{ \
\ \ }\alpha _{N-g,N-h}=\frac{1}{\sqrt{2}}\left( a_{g,h}-ib_{g,h}\right)
\nonumber \\
\alpha _{g,N-h} &=&\frac{1}{\sqrt{2}}\left( a_{g,N-h}+ib_{g,N-h}\right) ,%
\text{ \ \ \ }\alpha _{N-g,h}=\frac{1}{\sqrt{2}}\left(
a_{g,N-h}-ib_{g,N-h}\right)
\end{eqnarray}%
Here, $a_{g,h}$ and $b_{g,h}$ are independent Gaussian random numbers (real
numbers) with an average of zero and variance of one. Note that it is necessary to
generate $N\times N$ independent Gaussian random numbers for $N\times N$
fluctuations of $\eta _{g,h}$.

\begin{figure}[tbp]
\begin{center}
\includegraphics[scale=0.75]{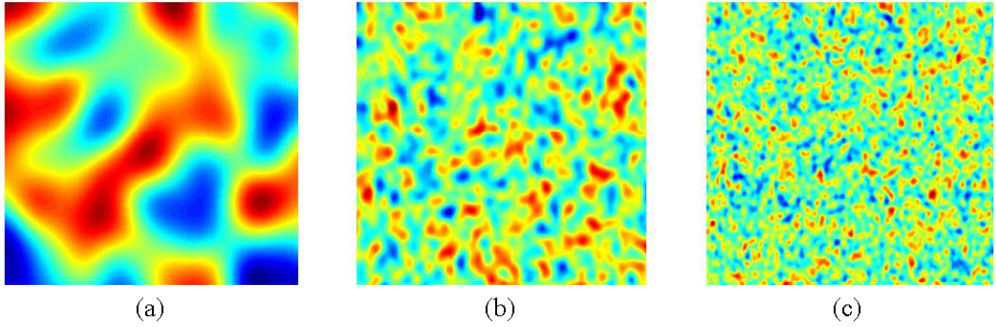}
\end{center}
\caption{The noise phase $\eta(x,z)$ with (a) correlation length $\tau=5.0\mu m$, (b)
correlation length $\tau=1.0\mu m$, and (c) correlation length $\tau=0.5\mu m$ in the studied
$48\mu m\times 40\mu m$ $xz$ plane.}
\label{noise}
\end{figure}

Finally, $\eta _{u,v}$ in
the discrete spatial space can be obtained straightforwardly by the Fourier
transform of $\eta _{g,h}$ in Equation \ref{disFT2}, such that the electric field $\mathbf{E}\left(
x,z\right) =\mathbf{E}_{0}\exp \left[ -i\mathbf{k}\cdot \mathbf{r}+\eta
\left( x,z\right) \right] $ in the studied $xz$ region can be evaluated. A further treatment of
$\eta _{u,v}\rightarrow \eta _{u,v}-\bar{\eta}$, in which $\bar{\eta}$ means the average noise over
the $N\times N$ grid data, is executed to ensure the condition
$\left\langle \eta \left( \mathbf{r}\right) \right\rangle=0$.
It is straightforward to extend to arbitrary $N\times M$ periodic grid data. Numerical results
of the noise phase with $\tau=5,1,0.5\mu m$ are illustrated in Figure \ref{noise} for reference.

\end{document}